\documentclass[conference]{IEEEtran}
\IEEEoverridecommandlockouts
\usepackage{cite}
\usepackage{amsmath,amssymb,amsfonts}
\usepackage{algorithmic}
\usepackage{graphicx}
\usepackage{textcomp}
\usepackage{xcolor}
\def\BibTeX{{\rm B\kern-.05em{\sc i\kern-.025em b}\kern-.08em
    T\kern-.1667em\lower.7ex\hbox{E}\kern-.125emX}}
\begin{document}

\title{LSTM-CNN Network for Audio Signature Analysis in Noisy Environments\\
}

\author{\IEEEauthorblockN{1\textsuperscript{st} Praveen Damacharla}
\IEEEauthorblockA{\textit{Research Scientist} \\
\textit{KineticAI Inc.}\\
The Woodlands, TX, USA \\
Praveen@KineticAI.com}
\and
\IEEEauthorblockN{2\textsuperscript{nd} Hamid Rajabalipanah}
\IEEEauthorblockA{\textit{Electrical Engineering} \\
\textit{Iran University of Science and Technology}\\
Tehran, Iran \\}
\and
\IEEEauthorblockN{3\textsuperscript{rd} Mohammad Hosein Fakheri}
\IEEEauthorblockA{\textit{Electrical Engineering} \\
\textit{Iran University of Science and Technology}\\
Tehran, Iran \\}
}

\maketitle

\begin{abstract}
There are multiple applications to automatically count people and specify their gender at work, exhibitions, malls, sales, and industrial usage. Although current speech detection methods are supposed to operate well, in most situations, in addition to genders, the number of current speakers is unknown and the classification methods are not suitable due to many possible classes. In this study, we focus on a long-short-term memory convolutional neural network (LSTM-CNN) to extract time and / or frequency-dependent features of the sound data to estimate the number / gender of simultaneous active speakers at each frame in noisy environments. Considering the maximum number of speakers as 10, we have utilized 19000 audio samples with diverse combinations of males, females, and background noise in public cities, industrial situations, malls, exhibitions, workplaces, and nature for learning purposes. This proof of concept shows promising performance with training/validation MSE values of about 0.019/0.017 in detecting count and gender.
\end{abstract}

\begin{IEEEkeywords}
Sound processing, artificial intelligence, deep learning
\end{IEEEkeywords}

\section{Introduction}
Presence detection is used for many applications. It can be used to automatically control the indoor climate, blinds or to inform staff on usage of certain rooms or office. Information on how many people are in a certain place is also useful for exhibition and product introduction to see which product attracts people and which gender\cite{b1, b2}. It is also useful for the visually or hearing impaired by giving them a better understanding. 
Many methods are used for people counting such as  Vision systems (cameras) in addition to WiFi and Bluetooth tracking are some of the most frequently used options. However, cameras are not placed everywhere, and WiFi tracking is unable to narrow a position down to a specific floor or room \cite{b3}. An alternative or additional solution which is considered in the following is to count people and specifying gender using sound. Using a variety of devices that are already present such as smartphones and economic recorders, sound can be recorded at minimal costs and difficulty.

\begin{figure}[!ht]
\centerline{\includegraphics[width=80mm]{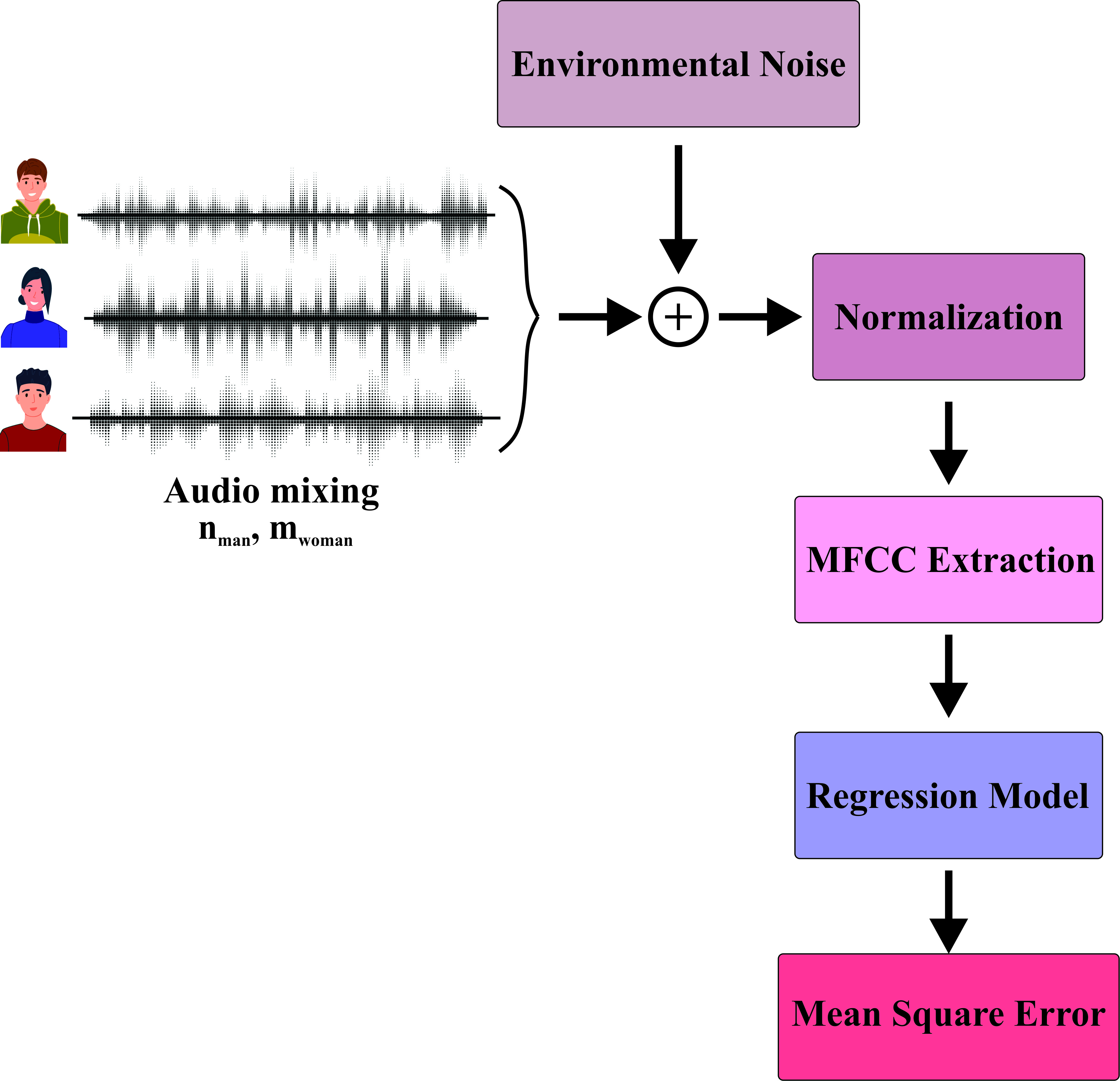}}
\caption{The illustration of the overall architecture for the proposed multiple speaker counting method.} 
\label{fig}
\end{figure}

\begin{figure*}[t]
\centerline{\includegraphics[width=180mm]{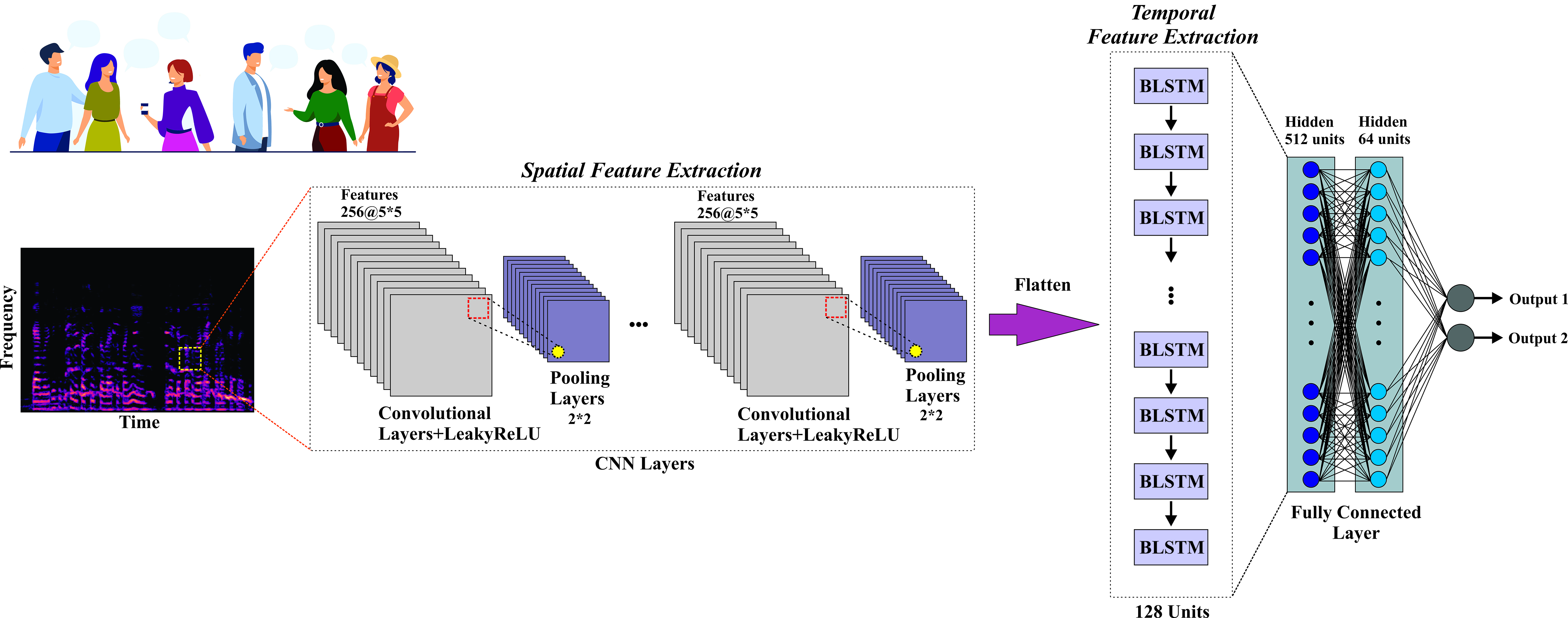}}
\caption{The proposed hybrid CNN-LSTM-FC architecture for estimating the gender/number of the speakers in noisy environments}
\label{fig}
\end{figure*}

\begin{figure}[!ht]
\centerline{\includegraphics[width=67mm]{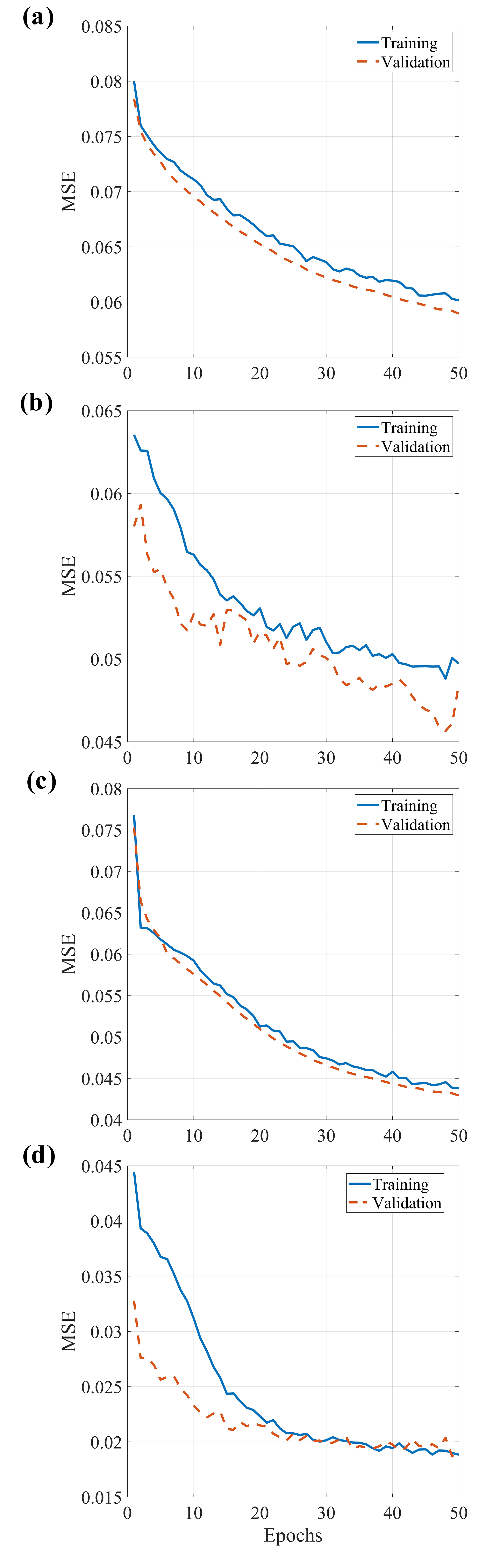}}
\caption{The MSE values corresponding to (a) FC, (b) CNN-FC, (c) LSTM-FC, and (d) CNN-LSTM-FC networks} 
\label{fig}
\end{figure}

Simultaneous conversations have been studied for several years yet still have many errors in many speech technology applications such as Speaker Identification and Automatic Speech Recognition \cite{b4, b5, b17, b18}. Many papers developed systems to recognize multiple speakers though the majority of them assume that the number of concurrent sources is known and also neglect gender and ambient noise \cite{b6, b7}, which is not a realistic assumption. Therefore, a semi-real-time, simple, yet effective speaker counts and gender estimation method is needed in noisy environment to become beneficial in real-world applications. Methods based on clustering also often require a priori information, e.g., the maximum number of concurrent sources in the processed sequence.

Artificial intelligence (AI) has made its entrance in the real-world applications in the last decades and is gaining considerable interest from the industrial community. Machine learning, Neural network and deep learning have become a popular data analysis tool across scientific society. Using machine learning with audio data has proven to be particularly useful in domains such as automatic speech recognition, \cite{b8} language processing, \cite{b9} and music classification \cite{b10} and event detection for urban and sport environments. 

Estimation of speaker-count usually needs two fundamental steps: (i) extracting the audio features related to the number of concurrent speakers, and (ii) classifying the number of speakers in the feature space. Recently, some attempts have been made to apply deep learning to audio source counting. In \cite{b11}, the convolutional neural network is used to classify the audio signals into 3 classes: 1, 2, and 3-or-more sources. Two main solutions may be found in the literature for this purpose, namely, regression and classification. Nevertheless, the clustering method is no longer efficient when both gender and number of active speakers are requested in a crowded environment. Although several studies have investigated speaker count estimation \cite{b4, b12, b13, b14}, they have not identified the gender of speakers. While, both gender/number data of customers may bring valuable information in various real-life industrial applications.  

This paper will focus on a regression-based deep learning approach to determine number and genders of people that are concurrently talking via the Long-Short-Term Memory Convolutional Neural Network (LSTM-CNN). We directly estimate a speaker count instead of counting them after identification. We have created 19000 samples of 5-s simultaneous speech with different combinations of male and female sounds. The contribution of this paper are summarized as follows:
\begin{itemize}
\item Beside estimating the count, the proposed algorithm is also able to detect the gender of speakers.
\item Unlike to previous proposals, the basis of our algorithm relies on regression model, which performs efficiently for gender detection in higher number of speakers.  
\item The proposed method retains its performance in noisy environments.
\end{itemize}

\section{Data Sets}
Many previous studies \cite{b15, b16}, have employed simulated overlapping speech for speaker counting with only 5-10\% of overlap speech. In this research, a big data set of noisy realistic conversation scenarios may happen in day life and industrial applications has been created. 
\subsection{Speech}
In order to train and test our network, a dataset containing concurrent speaking of different speakers with different gender, age, and accents. 3682 males and 2311 females have contributed to construct the dataset as speakers. For each sample, speaking sound of n males and m females (n+m$\le$10) are randomly merged. The dataset includes 19000 audio samples of concurrent talks each of which has 5-s duration and a sampling rate of 16000 frames per second where, silence in the beginning and the end of audios has been removed.

\subsection{Background Noise}
We include 50 non-speech (environmental sounds) examples in our training data, allowing network to better understand the possible background noises. The background sounds have been randomly selected among possibale real-life scenes such as industrial companies, work environments, exhibitions, public cities, malls, nature, farms, stores and etc., and then added to the original files as background noise (SNR$\approx$10 dB). 

\section{System Design}
Since the raw temporal waveforms are dense, it is not computationally efficient to use them, directly, as the input of the network model. Firstly, each of 19000 samples have been divided into 32000-frame overlapping samples with a 8000-frame shift. Through windowing the short-term signals and then applying the discrete Fourier transform, the Mel-frequency cepstral coefficients (MFCCs) of the raw data has been extracted, allowing us to use a compact representation of the spectrum as the input of the network model. 

\textbf{Fig. 1} displays the end-to-end procedure for gender/count detection based on audio signals. First, the speech voices of multiple persons ($\le$ 10) are randomly selected among the available dataset and then mixed. A randomly chosen environmental noise (SNR=$10$ dB) is added to the mixed signal. After normalizing the resultant signals, a time-windowing process is applied, making $q$-sample shorter signal. For each short-time window, the corresponding MFCCs are extracted to feed the regression model. The train-test splitting procedure is performed so as to allow the network see diverse combinations of men and women voices. Finally, the proposed regression network is learned by using the created dataset with two outputs, i.e., number of men and women in speakers.

Our proposed network architecture is depicted in \textbf{Fig. 2}, where the MFCCs are considered as the input data. The network is designed so as to extract both spatial and temporal features of the audio signals. The 2D convolutional network (2D-CNN) is responsible for extracting the spatial features and the temporal attributes are achieved by the bi-directional Long short-term memory (BLSTM) layer. Each CNN layer applies the linear operation by sliding the kernal of specified size on the output of the previous layer. The 2D-CNN overcomes the 1D-CNN for our application since it reduces the number of weights within a convolution layer through the parameter sharing concept. 

We use 7 CNN channels of kernal size 5$\times$5 with 256 filters in the network. Each convolution layer is followed by a Max-pooling layer with a dimension of 2$\times$2 to minimize the computational cost.  Padding is applied to keep the same dimensions after the convolutions. The Leaky Rectified Linear Unit (LeakyReLU, $\alpha=0.1$) function is used as the activation layer to increase the non-linearity degree and avoid the conventional dying ReLU problem. We should note that the audio signal have temporal correlation which should be investigated. Since the recurrent neural network (RNN) is unable to keep the information for the long sequences and suffers from small gradient updates, especially in the earlier layers, we use bidirectional Long short term memory (BLSTM, 128 units). Due to using both past and future information to update weights, the BLSTM network is more robust compared to the LSTM one. The feature maps extracted by CNN layers are then reshaped and then fed into 3 BLSTM layers. We will show that integrating the CNN with LSTM can significantly extract the spatial and temporal attributes of the raw data and therefore improve the overall accuracy of the speaker counting system.

To enhance the accuracy of regression, we employed a fully connected (FC) model with two hidden layers of 512 and 64 sizes. The dropout methods have been established to reduce the effect of overfitting and to enhance the capability of the detection model in unseen data. Also, the layers are initialized using Xavier Glorot's initialization. The final stage includes two nodes to reveal the number of men and women in the speakers. Evidently, the sum of these values determines the total number of speakers.

\section{Results}

We took care of using diverse combinations of man-woman speakers and environmental noises in the train, validation and test sets. The overall duration of the generated mixture dataset is 18 hours for training, and 5 hours for validation and test. The normalization pre-processing is applied to map the value of the features and also, the output values between 0 and 1. The train and test sizes are selected as 0.7 and 0.3, respectively.

\begin{table}
\caption{\label{demo-table}The MSE values resulted by different network models.}
\centering
\begin{tabular}{|c|c|c|c|c|}
\hline
Model &  FC  &  FC-CNN  &  FC-LSTM & FC-CNN-LSTM \\    \hline
MSE (Training) & 0.06 & 0.0497 & 0.0438 & 0.019\\    \hline
MSE (Validation) & 0.0589 & 0.0485 & 0.0429 & 0.017\\    \hline
\end{tabular}
\end{table}

To evaluate the performance of our regression problem, MSE performance indicator is employed to compare the exact value of the outputs with the observed ones. We investiage the MSE valuse for four different networks: 1) FC, 2) CNN-FC, 3) LSTM-FC, and 4) CNN-LSTM-FC to show that the proposed architecture has the minimum MSE value among all other combinations. The evaluation results are plotted in \textbf{Figs. 3a-d}, respectively. As expected, the FC model (four hidden layers of 1024, 512, 256, 64 nodes) has the highest MSE value of about 0.06 (\textbf{Fig. 3a}). Although the MSE values corresponding to CNN-FC (\textbf{Fig. 3b}) and LSTM-FC (\textbf{Fig. 3c}) are smaller, but they are not still convincing in our application. The simultaneous use of CNN and LSTM networks leads to the best training/validation MSE values of about 0.019/0.017 (\textbf{Fig. 3d}). The reason can be attributed to the fact that both spatial and temporal attributes of the short-frame audio signals are elaborately included in the learning process. The results are summarized in \textbf{Table I} which proves the effectiveness of the proposed hybrid model in the well-known speaker counting problem, even in the presence of the environmental noise.

\begin{table}
\caption{\label{demo-table}The MSE values of the proposed model corresponding to different Kernel sizes.}
\centering
\begin{tabular}{|c|c|c|c|}
\hline
 Kernel Size &  3$\times$3 &  5$\times$5 & 7$\times$7 \\    \hline
MSE (Training) & 0.0313 & 0.019 & 0.028 \\    \hline
MSE (Validation) & 0.0325 & 0.017 & 0.025 \\    \hline
\end{tabular}

\end{table}

\begin{table}
\caption{\label{demo-table}The MSE values of the proposed model corresponding to different CNN channel numbers.}
\centering
\begin{tabular}{|c|c|c|c|}
\hline
 CNN Channel Number & 3  &  5 &  7  \\    \hline
MSE (Training) & 0.0511 & 0.0311 & 0.019 \\    \hline
MSE (Validation) & 0.0482 & 0.0282 & 0.017 \\    \hline
\end{tabular}
\end{table}
\begin{table}
\caption{\label{demo-table}The MSE values of the proposed model corresponding to different CNN filter numbers.}
\centering
\begin{tabular}{|c|c|c|c|}
\hline
 CNN Filter Number &  64  & 128  &  256  \\    \hline
MSE (Training) & 0.039 & 0.0335 & 0.019 \\    \hline
MSE (Validation) & 0.037 & 0.0316 & 0.017 \\    \hline
\end{tabular}
\end{table}

\begin{figure}[!ht]
\centerline{\includegraphics[width=63mm]{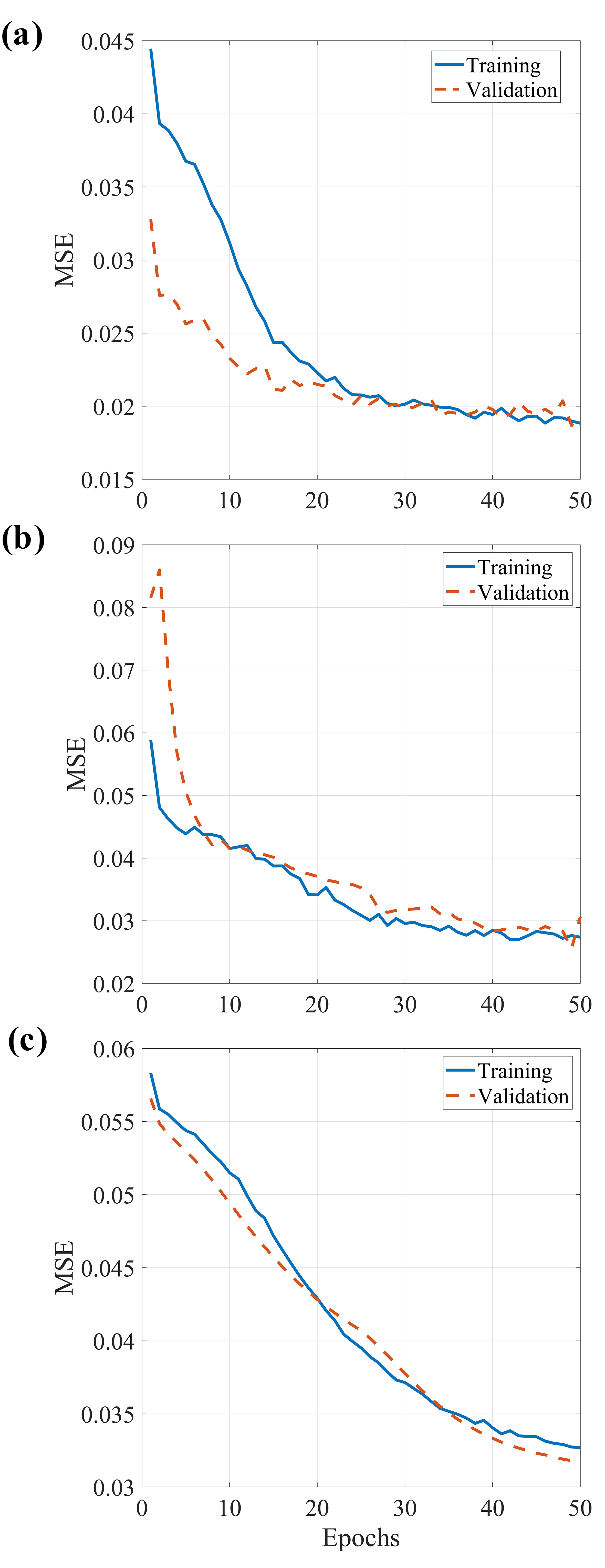}}
\caption{ Illustration of the effects of pre-windowing of the input audio signal on the MSE values of the proposed model. a) window size=q=32000 samples, shift=16000 samples, and b) window size=q=16000 samples,  shift=8000 samples, and c) window size=q= 8000 samples, shift= 4000 samples.} 
\label{fig}
\end{figure}

We should note that all contributing parameters of our proposed model have been through a comprehensive parametric study. For instance, we investigate the effect of change in the number of CNN filters, CNN channels and also, the kernel size. The corresponding MSE results are tabulated in \textbf{Tables II, III, IV}, respectively. Based on the results, we should remark the following points. As the CNN layers are responsible for extracting the complex features, the greater number of filters allowing the model to learn deeper features, yielding higher accuracy in gender/number estimation of speakers. As Table I shows, the model with 256 filters has the best MSE value. Using a greater number of filters may lead better response but at the expense of higher computing costs. Similar behaviors can be found in results of Table II, where considering the time-consumption, 7 layers are found high enough to attain acceptable MSE values. Finally, Table III reveals the optimum value of the CNN kernel size as which results in the training and validation MSE values of 0.019 and 0.017, respectively.

As mentioned before, the mixed audio signals have been divided to q-sample short-time windows feeding the network model. \textbf{Figs. 4a-c} depict the MSE values of the training system for different overlapped windows with q=8000, 16000, 32000 samples and q/2 shift corresponding to 0.5s, 1s, and 2s time duration (0.25 s, 0.5s, and 1s time shift), respectively. Evidently, a higher time span allows the model to deal with more information and learn better. The MSE values corresponding to 2-sec signals reach 0.019 (training) and 0.017 (validation), respectively. Therefore, the proposed model requires the audio input of at least 2-sec length to estimate well.

Additionally, we report the performance of our CNN-LSTM-FC architecture for different random sets of train-test data to show the robustness of the model. The network is trained by three different sets and the corresponding result are displayed in (\textbf{Fig. 5}). As can be seen, the MSE values of all three sets are close to 0.017, indicating that the proposed model is robust against the random splitting of the train-test data.

\begin{figure}[h]
\centerline{\includegraphics[width=120mm]{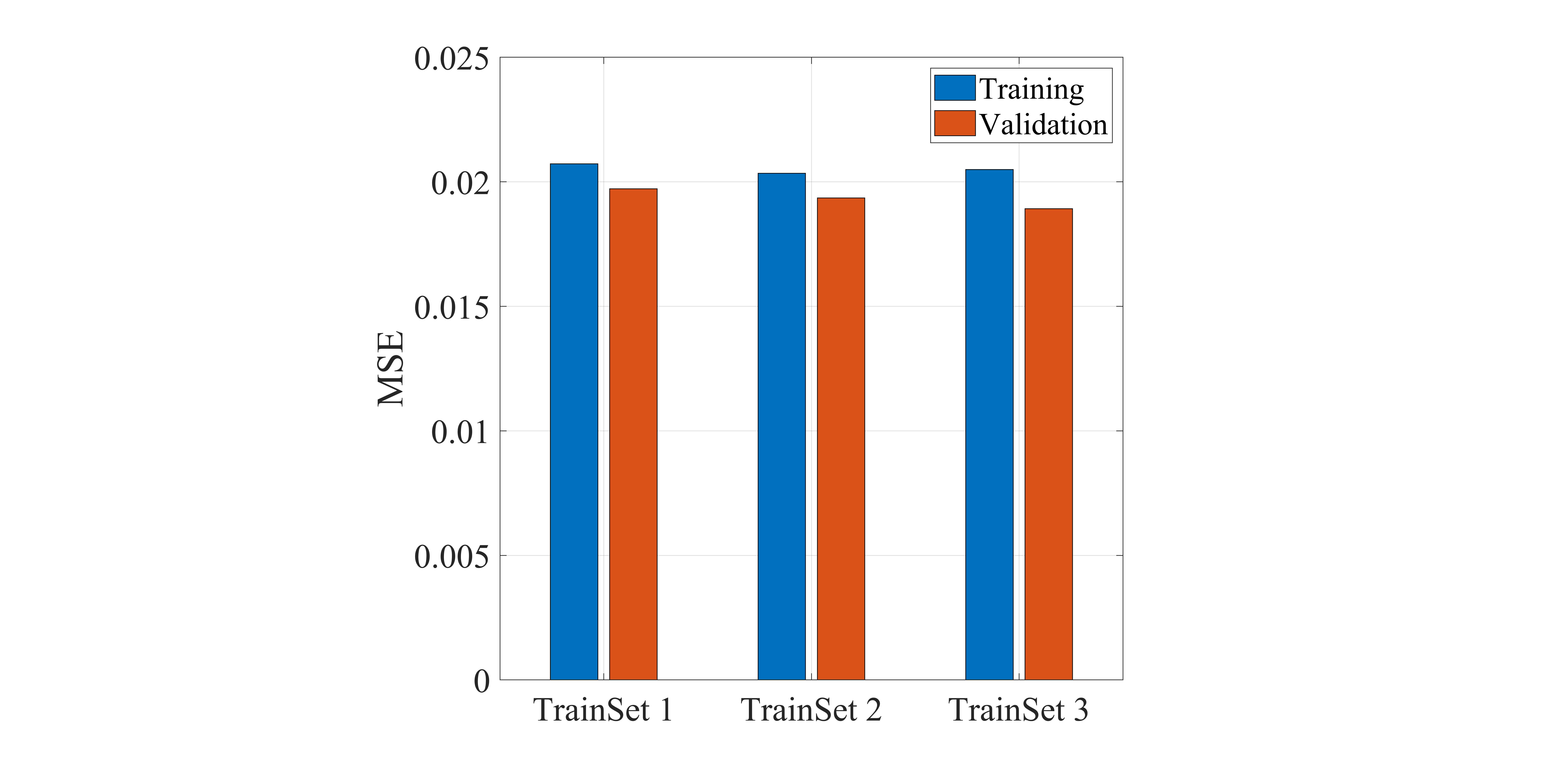}}
\caption{The MSE values of the proposed CNN-LSTM-FC architecture upon variation of train-test data splitting}
\label{fig}
\end{figure}

\section{Conclusions}
In this study, we proposed a hybrid CNN-LSTM-FC deep learning model for simultaneous detection of gender/number of speakers in multi-speaker noisy environments. The CNN and LSTM layers were used to extract the spatial and temporal attributes of the speech, respectively. The dataset includes 19000 audio samples of concurrent talks each of which has 5-s duration and a sampling rate of 16000 frames per second where, silence in the beginning and the end of audios has been removed. A comprehensive parametric study was carried out to find the best set of contributing parameters in the training model.  The proposed network has been trained and our results showed the MSE value of about 0.017 in estimating the number of men and women between the speakers. The MSE value has been reported for different network architectures to prove that the effectiveness of the proposed model overcomes the other possible solutions.

\vspace{12pt}
\color{red}

\end{document}